\documentclass[fleqn,usenatbib]{mnras}

\usepackage[T1]{fontenc}
\usepackage{ae,aecompl}

\usepackage{siunitx}
\usepackage{graphicx}
\title[X-rays from candidate Lyman emitting galaxies]{Enhanced X-ray emission from candidate Lyman continuum emitting galaxies}

\author[Bluem, Kaaret, Prestwich, Brorby ]{
J. Bluem$^{1}$\thanks{E-mail: jesse-bluem@uiowa.edu},
P. Kaaret$^{1}$ \thanks{E-mail: philip-kaaret@uiowa.edu},
A. Prestwich$^{2}$,
M. Brorby$^{2}$
\\
$^{1}$Department of Physics and Astronomy, University of Iowa, Iowa City, IA 52245, USA\\
$^{2}$Harvard-Smithsonian Center for Astrophysics, 60 Garden Street, Cambridge, MA 02138, USA\\}

\date{Accepted XXX. Received YYY; in original form ZZZ}

\pubyear{2018}

\begin{document}
\label{firstpage}
\pagerange{\pageref{firstpage}--\pageref{lastpage}}
\maketitle

\begin{abstract}
X-ray binaries may have helped reionize the early Universe by enabling Lyman continuum escape. We analyzed a set of 8 local galaxies that are potential Lyman leaking galaxies, identified by a blue color and weak emission lines, using {\it Chandra} X-ray observations. Five of the galaxies feature X-ray sources, while three galaxies are not significantly detected in X-rays. X-ray luminosities were found for the galaxies and X-ray sources. Four of the galaxies have elevated X-ray luminosity versus what would be expected based on star formation rate and metallicity. The presence of detected X-ray sources within the galaxies is found to correlate with the ratio of the star formation rate estimated from the near-ultraviolet flux to that estimated from the infrared. This implies reduced obscuration due to dust in the galaxies with X-ray sources. These results support the idea that X-ray binaries may be an important part of the process of reionziation.
\end{abstract}


%%%%%%%%%%%%%%%%% BODY OF PAPER %%%%%%%%%%%%%%%%%

\section{Introduction}

There are many unanswered questions about the early universe. One such question is the mechanism by which the universe switched from being cool and neutral to hot and ionized \citep{Mesinger2013}. This period of time is often referred to as the epoch of reionization and occurred during a time frame corresponding to a mean redshift of $\rm  11.0 \pm 1.4$ \citep{Dunkley2009}, and completed at roughly z = 6 \citep{Fan2006, Bosman2018} (roughly 400 Myr and 900 Myr, respectively). The bulk properties of the universe at this time differed in significant ways from the current universe, beyond the ionization fraction of the intergalactic medium (IGM). Evidence shows that the galactic radiation escape fraction has decreased over time \citep{Inoue2006, Siana2010} and metallicity was lower in early galaxies \citep{Basuz2013,Carniani2018}. During the epoch of reionization, star formation rates and X-ray luminosities were higher in galaxies \citep{Basuz2013} and dwarf galaxies were more common \citep{Alavi2016}. These are all factors that may affect the reionization of the universe. 

The ionization of neutral hydrogen requires ultraviolet (UV) photons in the Lyman continuum, corresponding to the extreme end of the UV spectrum. The source of such UV photons is most commonly thought to be massive stars in early galaxies, which produce plentiful UV radiation \citep{Loeb2010, Heckman2011}, but it is also possible that the very first (population III) stars or accreting black holes contributed as well \citep{Madau2004,Mirabel2011}.  However, UV radiation is readily absorbed by gas and dust in the star's host galaxy \citep{Wofford2013}. If these UV photons do not escape their parent galaxy, then they cannot ionize the IGM. Looking to the average local galaxy, this is exactly the situation observed - minimal radiation escapes the galaxy \citep{Leitherer1995, Heckman2011, Wofford2013}. To assist in the escape of these photons, the effect of dust and gas in the galaxy must then be reduced in some way \citep{Orsi2012, Wofford2013}. This reduction in dust and neutral gas could apply to the entire galaxy, or only to select localized paths out of the galaxy (the picket-fence model). Observations of partial ionization in the known Lyman leaker Haro 11 support the picket-fence model \citep{Keenan2017}. 

Modeling of galaxies shows that supernova outflows enhance the capability of the host galaxy to leak Lyman radiation \citep{Orsi2012}. Observations further support that outflows may be a defining feature of leaking galaxies, in part due to the lack of any other consistent feature when compared to non-leakers \citep{Wofford2013}. Extensive outflows of wind-blown material has been detected in local analogs to early galaxies, and have proven to be a common feature for these galaxies \citep{Shapley2003}. Jets and accretion-induced winds from compact objects may also contribute to clearing dust and ionizing neutral gas, including feedback from X-ray binaries (XRBs) \citep{Prestwich2015}. XRBs are multiple star systems consisting of a compact object (black hole or neutron star) and a normal star \citep{Orosz2011}. If the companion is a massive star, then it is considered a high mass XRB (HMXRB). Observations show that the power contained in jets from black holes can potentially match or exceed the radiative luminosity of the compact object, by up to a few orders of magnitude \citep{Gallo2005, Pakull2010}. \Citet{Pakull2010} calculated the mechanical power of a black hole as $\rm 5 \times 10^{40}$ $\rm erg$ $\rm s^{-1}$, exceeding the object's X-ray luminosity by a factor of $\rm 10^4$. For comparison, the brightest X-ray binaries are observed to reach luminosities in excess of $10^{41}$ $\rm erg$ $\rm s^{-1}$ \citep[for a review see][]{KaaretFeng2017}.

High mass XRBs are affiliated with the same young stellar associations that contain the aforementioned massive stars, resulting in more XRBs when there is more star formation \citep{Grimm2003, Ranalli2003, Kaaret2008, Mineo2012}.  It is this relationship that results in star formation rate (SFR) correlating with X-ray luminosity. Observational evidence shows that XRB populations increase with lower metallicity \citep{Brorby2014, Prestwich2013, Kaaret2011}; this suggests that XRBs were more common in the early universe. \citet{Basuz2013} directly observed the effect of elevated X-ray luminosity versus SFR in galaxies at high redshift, when metallicity was lower. \Citet{Basu2013} also observed a similar effect in low metallicity, low redshift galaxies. This observed X-ray excess implies more XRBs than expected based on scaling from nearby, near-solar metallicity galaxies.

Another factor that can help with creating channels out of a galaxy in the picket-fence model is distorting the host galaxy through mergers \citep{Bergvall2013}. This can result in galaxies with tadpole style morphology. These Tadpole galaxies are more common at high redshift values and feature elevated SFRs \citep{Elmegreen2007,Elmegreen2012}. Mergers can result in off-center star formation regions, effectively moving the UV generating massive stars and XRBs to the edge of the galaxy, reducing column density and encouraging leakage \citep{Bergvall2013}. The local Lyman leaking galaxy Haro 11 shows evidence of a past merger \citep{Ostlin2015}. Ram pressure stripping can also reduce gas density on the leading edge of a galaxy as it moves through the IGM, resulting in ``bare" stars that have a favorable environment for leaking ionizing radiation to the IGM \citep{Kronberger2008}. During the process of this stripping, SFR can increase as well \citep{Kronberger2008}. Both of these effects combined lead to another favorable scenario for leakage.

Unfortunately, high redshift observations of the sort needed to test these ideas are difficult to acquire. The available instruments generally lack the resolution needed, and the observation times required are unfeasible \citep{Basu2013}. The Lyman alpha forest is too dense when observing the early universe to identify Lyman leakage \citep{Kim2007}, and the IGM is also highly opaque at that time \citep{Gunn1965, Bosman2018}. Often in order to get sufficient signal-to-noise to study galaxies from this ancient era, the observations of different galaxies must be stacked, resulting in studying the general properties of galaxies, and not the particulars of any single galaxy \citep{Nandra2002, Basuz2013}. This makes understanding the factors behind Lyman leakage difficult. A solution to these observational problems is to study local galaxies that have features analogous to these high redshift galaxies.

Certain parameters must be met for these analog galaxies to match the conditions of the high redshift galaxies. The analog galaxies need to feature low metallicities and have young starbursts, with ages of 100 million years or less, so that the stars in the local galaxy are of a similar age to those in the early universe \citep{Bergvall2013}. Disturbed morphology is good for matching early galaxies, but not necessarily a requirement. It is through studying these analogs that we hope to gain insight into the processes at play in the high red-shift universe. However, there are problems with this; as discussed earlier, some of the parameters of the universe have changed since then.

In particular, thus far observed escape fractions are far too low to account for the level of ionizing radiation required for the reionization of the universe. An escape fraction of roughly 20\% is expected based on results from the Wilkinson Microwave Anisotropy Probe (WMAP) \Citep{Bouwens2012}. \Citet{Leitherer1995} finds an escape fraction of less than 3\% in their sample of galaxies and \Citet{Leitet2011} finds a similarly low escape fraction for local analog Haro 11 of $\rm 3.3 \pm 0.7$\%. The recent results from \Citet{Puschnig2017} lower the escape fraction for Tololo 1247-232 to $\rm 1.5 \pm 0.5$\%. \Citet{Deharveng2001} finds an upper limit to the escape fraction of 6.2\% for Mrk 54. \citet{Bergvall2013} suggests that perhaps the reason for such low observed escape fractions is that the criteria for selecting local analogs is faulty. 

\citet{Bergvall2013} proposed new criteria for selecting potential Lyman continuum leaking galaxies: weak emission lines and a strongly blue color (u-g < 0.7). Blue color implies young, giant stars - which provide sufficient UV radiation for ionizing hydrogen. This population of short-lived massive stars also provides XRBs. The weak emission lines are the opposite of the previous convention for target criteria. The logic behind strong H$\rm\alpha$ emission lines is that they serve to identify that the radiation required to ionize hydrogen is in fact being produced. However, the problem with this is that strong emission correlates with high column densities. As such, the amount of ionization occurring means that the radiation may not be escaping the galaxy, and is instead ``consumed" by the high column density. By looking for weak emission this problem is effectively dodged, since if the gas is already mostly ionized or the column density is low, we wouldn't expect strong emission lines. In effect, this method emphasizes looking for the conditions favoring leakage rather than looking for the conditions favoring UV radiation production. This paper studies the X-ray properties of the eight strongest candidate analog galaxies found using this method from \citet{Bergvall2013}, using imaging with the Chandra X-ray Observatory. 

\section{Sample}\label{observations}
\label{sec:obs}

Our sample consists of eight blue compact galaxies. These eight galaxies are potential Lyman continuum leakers described in \citet{Bergvall2013}. \citet{Bergvall2013} filtered out galaxies hosting an AGN when selecting the sample. Table 1 includes measured parameters for these eight galaxies. Tadpole morphology is seen in three galaxies in our sample, and six galaxies in total show very clear signs of perturbation by merger \citep{Bergvall2013}. While Galex J010724.7+13320 appears spherical in visual images, it has a chaotic appearance in H$\rm \alpha$ \citep{Bergvall2013}. Overall, seven of the galaxies show evidence of mergers or interaction with neighbors \citep{Bergvall2013}. Table 1 includes information about the morphology of these galaxies.

%===TABLE=====================================================================================
\begin{table*}
%\begin{minipage}{150mm}
\caption{Candidate Lyman continuum leaking galaxies.  The table includes for each galaxy: the galaxy name, reference number, distance, $\rm n_H$ column density, metallicity, and geometry (as described by \citet{Bergvall2013}. A 'T' in the geometry column specifies a tadpole galaxy, and a 'P' specifies a galaxy that appears perturbed. Galaxy reference number is maintained for each galaxy in all data tables.}
\centering
\begin{tabular}{lccccc}
\hline
Galaxy name                    & Number & Distance & $\rm n_H$ & Metallicity & Morphology\\
						&&	(Mpc)		& ($\rm cm^{-2}$)		& (12+log(O/H))&\\
\hline
GALEX J010724.7$+$133209 &1& 167.5  & $3.28 \times 10^{20}$ & 8.38   & \\
GALEX J024352.8$-$003703 & 2 & 131.2    & $2.96 \times 10^{20}$& 10.2   & P\\
GALEX J025325.8$-$001357 & 3 & 118.0    & $5.54 \times 10^{20}$ & 8.42  & P\\
GALEX J075313.3$+$123749 &4& 128.2   & $3.12 \times 10^{20}$ & 8.11  & P, T\\
GALEX J080754.6$+$141045 & 5 & 127.1    & $2.92 \times 10^{20}$ & 8.34   & P, T \\
GALEX J085642.1$+$123157 &6& 130.6   & $3.49 \times 10^{20}$ & 8.06 & P\\
GALEX J100712.2$+$065736 & 7 & 138.0  & $1.9 \times 10^{20}$ & 8.11  & P, T\\
GALEX J101007.7$+$033130 &8& 140.0   & $2.07 \times 10^{20}$ & 8.14   & \\
\hline
\end{tabular}
\\
\raggedright
\vspace{1mm}
\textbf{Notes.} Distances are from HyperLeda. $\rm n_H$ column densities are from HEASARC. Metallicities are the ON method metallicities from \citet{Bergvall2013}, using the method described in \citet{Pilyugin2010}.\\
\end{table*}
%=============================================================================================

\section{Galactic D25 Ellipses and Star Formation Rates}\label{results}
\label{sec:results}
%\section{Discussion}\label{discussion}
%\label{sec:discussion}

D25 ellipses were manually found using SDSS g-band visual images (rather than the traditional B-band, as there are no B-band images on SDSS). A contour corresponding to 25th magnitude per square arcsecond \citep{deVaucouleurs1991}, after converting to nanomaggies, was drawn on each visual image. Starting with central coordinate information and position angles from HyperLeda, the right ascension and declination, as well as position angles and central coordinates were adjusted to match the contour. Overall, the new D25 ellipses have significantly smaller radii when compared to the HyperLeda D25 ellipse values. The apparent sizes of the galaxies range from 8\arcsec\ $\rm \times$ 7\arcsec\ to 20\arcsec\ $\rm \times$ 12\arcsec.

SFR estimates (in solar masses per year) were done following the procedure outlined in \citet{Mineo2012}. The IR component of SFR was found using WISE band 4 magnitudes, converted to monochromatic luminosities as described in \citet{Wright2010}. These monochromatic luminosities were then used with the \citet{Chary2001} SED templates to get the 8-1000 micron band infrared (IR) luminosity. This luminosity was then used with the following equation to get a star formation rate, as per \citet{Mineo2012}:
\begin{equation}
\rm SFR_{IR} = 4.6 \times 10^{-44} \: L_{IR}(erg \: s^{-1})
\end{equation}
Near ultraviolet (NUV) luminosity was calculated using NUV images from GalexView. Luminosity is found using a counts to flux conversion factor from GALEX of $2.06\times10^{-16}$ $\rm erg$ $\rm cm^{-2}$ $\rm s^{-1}$ $\rm \si{\angstrom}^{-1}$ / $\rm ph$ $\rm s^{-1}$ and the count rates from the redefined D25 ellipses described earlier, along with the distances to the galaxies listed in Table 1. These fluxes were then converted to luminosities and used in the following star formation rate formula from \citet{Mineo2012}:
\begin{equation}
\rm SFR_{NUV} = 1.2 \times 10^{-43} \: L_{NUV}(erg \: s^{-1})
\end{equation}
A total SFR can be calculated from the IR and UV SFRs \citep{Mineo2012}:
\begin{equation}
\rm SFR_{tot} = SFR_{NUV} + (1-\eta)SFR_{IR}
\end{equation}
 $\rm SFR_{tot}$ is the total star formation rate. The correction factor $\eta$ accounts for the portion of IR luminosity from aged stars versus young stars, and is taken to be zero given that all galaxies in the sample are assumed to be starburst galaxies \citep{Hirashita2003}. SFR results are in Table 2.
 
%===TABLE=====================================================================================
\begin{table*}
%\begin{minipage}{150mm}
\caption{SFR data for candidate Lyman continuum leaking galaxies. The table includes for each galaxy: assigned number, luminosities in NUV and IR (IR luminosity is directly converted from WISE band 4 catalog magnitudes), calculated star formation rates in NUV and IR, as well as the total star formation rate as a combination of both.}
\centering
\begin{tabular}{lccccc}
\hline
Galaxy            & NUV luminosity& IR luminosity&$\rm SFR_{NUV}$     & $\rm SFR_{IR}$ & $\rm SFR_{TOTAL}$           \\
                      &$\rm (erg s^{-1})$&$\rm (erg s^{-1})$ &($\rm M_{\sun} $ $\rm yr^{-1}$)&  ($\rm M_{\sun} $ $\rm yr^{-1}$)      & ($\rm M_{\sun} $ $\rm yr^{-1}$)       \\ \hline 
1 &$1.01 \times 10^{43}$&$4.54 \times 10^{42}$& 1.21 & 1.66& 2.87 \\
2&$5.85 \times 10^{42}$&$4.62 \times 10^{41}$& 0.70 & 0.20& 0.90 \\
3&$1.16 \times 10^{43}$&$2.28 \times 10^{42}$& 1.39 & 0.86& 2.25 \\
4 &$2.03 \times 10^{42}$&$1.07 \times 10^{42}$& 0.24 & 0.44& 0.68\\  
5 &$8.73 \times 10^{42}$&$1.63 \times 10^{42}$& 1.05 & 0.63&   1.68\\
6 &$5.76 \times 10^{42}$&$1.33 \times 10^{42}$& 0.69 & 0.54& 1.23  \\
7 &$9.31 \times 10^{42}$&$2.17 \times 10^{42}$& 1.12 & 0.82& 1.94 \\ 
8 &$3.89 \times 10^{42}$&$7.00 \times 10^{41}$& 0.47 & 0.29& 0.76 \\                           
\hline
\end{tabular}
\\
\raggedright
\vspace{1mm}
\end{table*}
%============================================================================================= 

\section{X-ray Observations and Analysis}\label{xray}
\label{sec:xra}

The galaxies were observed with Chandra, in two batches. The initial four galaxies in the sample are sequential observation IDs 16972-16975. Those observations took place in November and December 2014, and June 2015. The second batch of four are sequential observation IDs 19327-19330, with observations in November and December 2016, as well as October and December 2017. Figures 1 and 2 show the X-ray images for each galaxy, with the redrawn D25 ellipses and all detected X-ray sources marked.

\begin{figure*}
\centering
\includegraphics{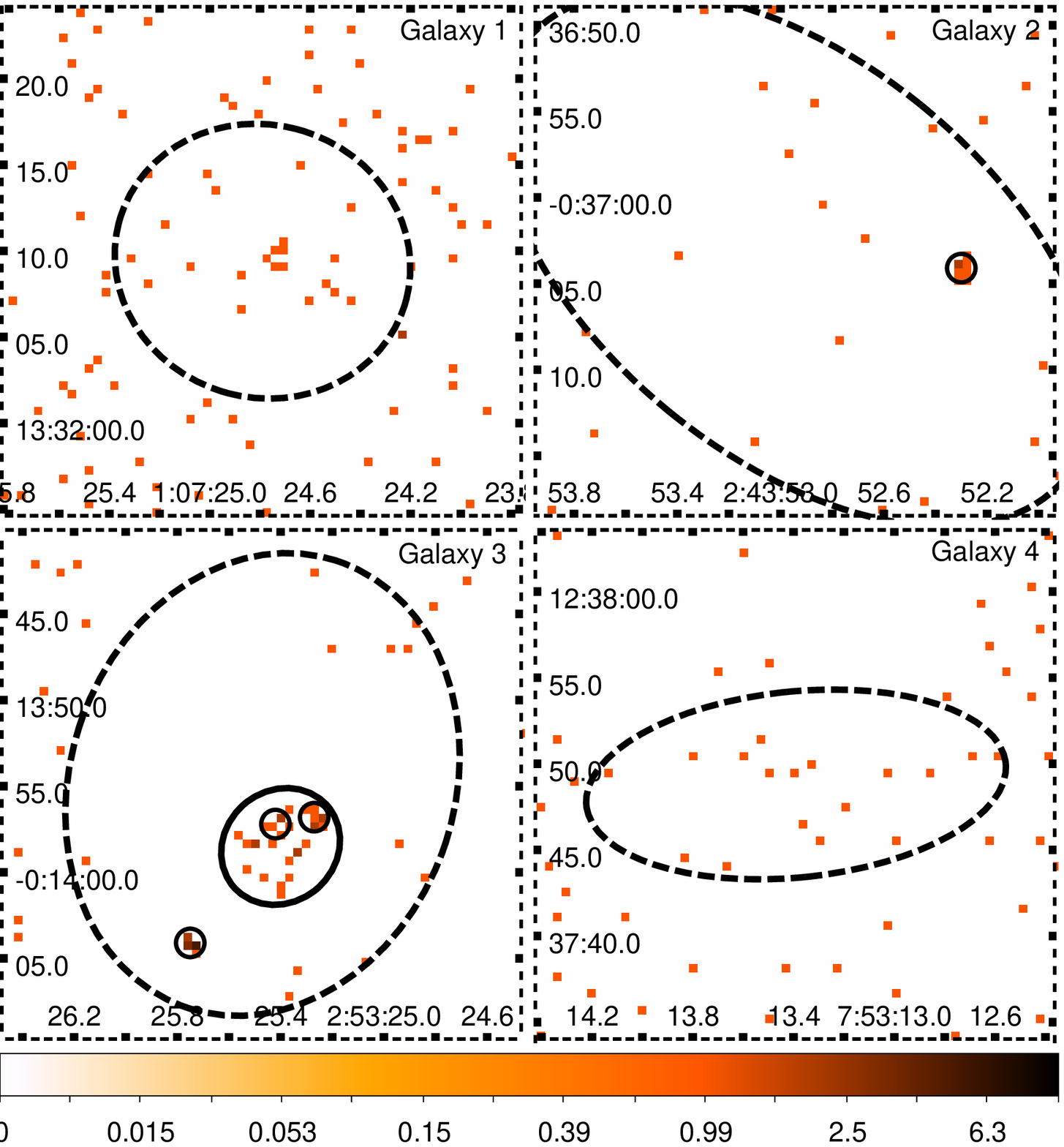}
\caption{X-ray images for galaxy 1-4. The large dashed ellipse in each image is the D25 ellipse for the galaxy. Each identified point source is labeled with a circle showing the area analyzed with {\tt srcflux}. The small solid ellipse in the GALEX J025325.8-001357 (Galaxy 3) image is the region of extended emission. Axes are right ascension and declination. North is up. Color bar units are counts.}
\end{figure*}

\begin{figure*}
\centering
\includegraphics{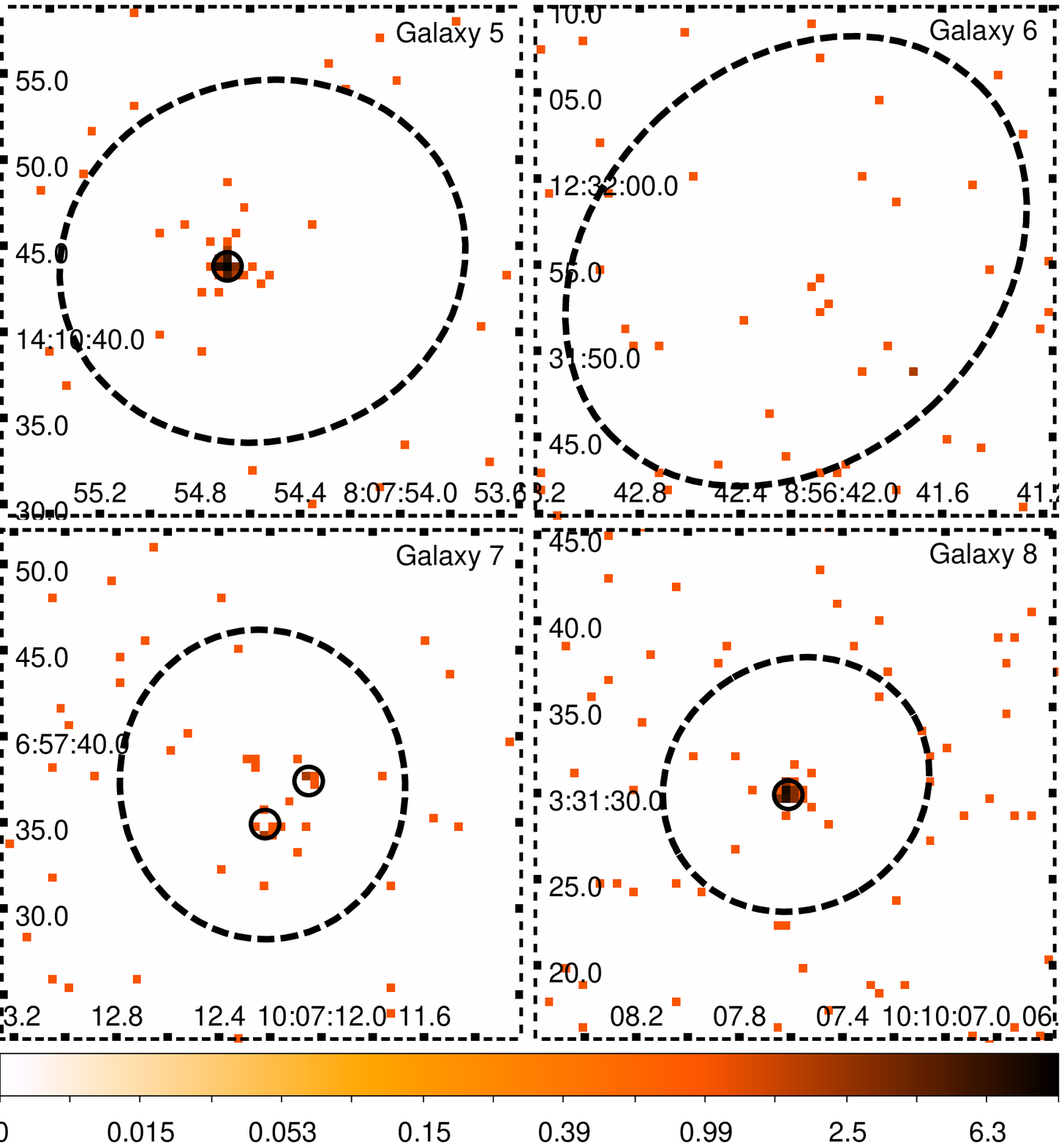}
\caption{X-ray images for galaxy 5-8. The large dashed ellipse in each image is the D25 ellipse for the galaxy. Each identified point source is labeled with a circle showing the area analyzed with {\tt srcflux}. Axes are right ascension and declination. North is up. Color bar units are counts.}
\end{figure*}

All X-ray analysis was handled with CIAO 4.9 (Chandra Interactive Analysis of Observations), Chandra's data analysis system \citep{Fruscione2006}. The CALDB version (calibration files) used was 4.7.7. The Chandra images of the eight galaxies were first processed using the {\tt chandra\_repro} tool in CIAO. Exposure corrected images and exposure maps were made with the {\tt fluximage} tool and PSF (point spread function) maps were made with the {\tt mkpsfmap} tool. The tools {\tt wavdetect} and {\tt celldetect} were used to identify and confirm sources, with {\tt wavdetect} identifying point sources and regions of extended emission, and with {\tt celldetect} serving to verify the point sources identified with {\tt wavdetect}. 
 
For each source and galaxy in the sample, X-ray fluxes were calculated from the net counts in the 0.3-8 keV range, then converted to luminosities using the Mpc distances from Table 1. The source flux is found by using a central coordinate for the point source. In the case of discrete point sources,  the central coordinate was simply the coordinates returned with {\tt wavdetect}. However, for the extended emission regions identified by {\tt wavdetect}, areas of high counts within the extended emission region were identified manually and their central coordinates run as point sources. Three such point sources were identified within the two regions of extended emission in our sample. Total galactic X-ray flux was found using our D25 ellipses as the source region. Background regions were specified in the image, as an annulus around the source region, making sure that no detected X-ray sources (as found by {\tt wavdetect}) infringe on the background annulus. All source and galaxy fluxes were found using the CIAO tool {\tt srcflux}, returning an unabsorbed flux in the 0.3-8 keV band, using an absorbed power-law model. A photon index of 1.5 was used, along with appropriate $\rm n _H$ values from the High Energy Astrophysics Science Archive Research Center (HEASARC) for each galaxy (sources were run with a $\rm n _H$ column density matching that of the host galaxy). Galaxies were run with a psfmethod setting of ``ideal" in {\tt srcflux} (as is appropriate for an extended region), with D25 ellipses as the source regions and background annuli for background subtractions, while point sources in the galaxies were run with a psfmethod of ``arfcorr", simply specifying the source coordinates.

%===TABLE=====================================================================================
\begin{table*}
%\begin{minipage}{150mm}
\caption{Total X-ray luminosities for candidate Lyman continuum emitting galaxies and sources within the respective galaxies. The table includes: target names for the respective galaxy (sources labeled with the CXOU prefix for Chandra), Chandra exposure time for the galaxy, net background subtracted counts for the region, total X-ray flux and luminosity for the region, and predicted total X-ray luminosity, estimated from total SFR using equation 22 from \citet{Mineo2012}, and the Poisson probability of a source being spurious. Also included is the extended emission region in galaxy 3, and summed sources for galaxies featuring multiple sources.}
\centering
\renewcommand{\arraystretch}{1.2}
\begin{tabular}{lcccccc}
\hline

Galaxy or region & Exposure & Net counts & X-ray flux           & X-ray luminosity & $\rm L_X$ from SFR & Probability\\
                       &ks &        & (erg $\rm cm^{-2} $ $\rm s^{-1}$)       & (erg $\rm s^{-1}$) & (erg $\rm s^{-1}$) & \\ \hline 
Galaxy 1&35.6&&&&$7.49 \times 10^{39}$\\
D25 ellipse& & 2.8 $\pm$ 4.8& <$3.75 \times 10^{-15}$ & <$1.26 \times 10^{40}$ & &\\
\\
Galaxy 2&17.5&&&&$2.35 \times 10^{39}$\\
D25 ellipse&  & 1.8 $\pm$ 5.1& <$6.57 \times 10^{-15}$ & <$1.35 \times 10^{40}$ &  &\\
CXOU J024352.3$-$003704 && 7.0 $\pm$ 2.7 & $4.70^{+3.58}_{-2.37} \times 10^{-15}$ & $9.68^{+7.37}_{-4.88} \times 10^{39}$&&$5.4 \times 10^{-6}$\\
\\
Galaxy 3&14.5&&&&$5.87 \times 10^{39}$\\
D25 ellipse &  & 37.6 $\pm$ 7.2& $2.77^{+0.95}_{-0.81} \times 10^{-14}$ & $4.61^{+1.58}_{-1.35} \times 10^{40}$ & &\\
CXOU J025325.8$-$001404 &  & 11.0 $\pm$ 3.3 & $8.35^{+4.95}_{-3.51} \times 10^{-15}$ & $1.39^{+0.83}_{-0.59} \times 10^{40}$&&$1.0 \times 10^{-12}$\\
CXOU J025325.3$-$001357 &   & 7.4 $\pm$ 2.8 &$6.29^{+4.90}_{-3.28} \times 10^{-15}$ & $1.05^{+0.82}_{-0.55} \times 10^{40}$&&$6.5 \times 10^{-8}$\\
CXOU J025325.4$-$001357 &   & 5.1 $\pm$ 2.5 & $4.31^{+4.38}_{-2.77} \times 10^{-15}$ & $7.18^{+7.30}_{-4.61} \times 10^{39}$&&$5.8\times 10^{-5}$\\
Extended emission region&&28.0 $\pm$ 5.4 & $2.06^{+0.73}_{-0.59} \times 10^{-14}$ & $3.43^{+1.22}_{-0.98} \times 10^{40}$&&\\
Sum of sources&&&$2.90^{+0.88}_{-0.69} \times 10^{-14}$&$4.82^{+1.48}_{-1.14} \times 10^{40}$\\
\\
Galaxy 4&20.8&&&&$1.77 \times 10^{39}$\\
D25 ellipse&  & 5.9 $\pm$ 4.1& <$7.56 \times 10^{-15}$ & <$1.49 \times 10^{40}$ &  &\\   
\\
Galaxy 5&16.2&&&&$4.38 \times 10^{39}$\\
D25 ellipse&  & 58.7 $\pm$ 8.4&   $3.90 \pm 0.92 \times 10^{-14}$ & $7.54 \pm 1.78 \times 10^{40}$ & &\\
CXOU J080754.7$+$141044 & & 51.5 $\pm$ 7.2& $3.98 \pm 0.92 \times 10^{-14}$ & $7.69 \pm 1.78 \times 10^{40}$&&$3.3 \times 10^{-97}$\\
\\
Galaxy 6&22.4&&&&$3.21 \times 10^{39}$\\
D25 ellipse&  & -6.2 $\pm$ 5.1& <$4.19 \times 10^{-15}$ & <$8.55 \times 10^{39}$ & &\\
\\
Galaxy 7&19.2&&&&$5.06 \times 10^{39}$\\
D25 ellipse&  & 16.4 $\pm$ 4.9& $9.39^{+9.38}_{-4.17} \times 10^{-15}$ & $2.14^{+2.14}_{-0.95} \times 10^{40}$ & &\\      
CXOU J100712.2$+$065735 && 4.4 $\pm$ 2.2 & $2.93^{+3.16}_{-1.93} \times 10^{-15}$ & $6.68^{+7.20}_{-4.40} \times 10^{39}$&&$1.4 \times 10^{-3}$\\
CXOU J100712.1$+$065737       & & 3.6 $\pm$ 2.0 & $2.36^{+2.90}_{-1.67} \times 10^{-15}$ & $5.38^{+6.61}_{-3.80} \times 10^{39}$ &&$2.4\times 10^{-2}$\\
Sum of sources&&&$5.29^{+4.29}_{-2.55}\times 10^{-15}$&$1.21^{+0.98}_{-0.58} \times 10^{40}$\\
\\
Galaxy 8&26.1&&&&$1.98 \times 10^{39}$\\
D25 ellipse&  & 31.4 $\pm$ 6.6& $1.57^{+0.59}_{-0.50} \times 10^{-14}$ & $3.68^{+1.38}_{-1.17} \times 10^{40}$ &  &\\     
CXOU J101007.6$+$033130 8& & 28.6 $\pm$ 5.4& $1.72^{+0.58}_{-0.47} \times 10^{-14}$ & $4.03^{+1.36}_{-1.10} \times 10^{40}$&&$6.1 \times 10^{-37}$\\
\hline
\end{tabular}
\\
\raggedright
\vspace{1mm}
\end{table*}
%============================================================================================= 

Table 3 includes observation information, measured flux, and luminosity values for all eight galaxies and the sources detected within the galaxies. Eight sources were found in five of the eight galaxies in the sample, with one galaxy hosting three sources and one galaxy hosting two sources. The other three galaxies are not detected significantly in X-rays. Also included in Table 3 are the sums of the X-ray luminosities of the sources within each galaxy, and the probability of the detected sources being spurious. In subsequent analysis, the sum of source luminosities is used as a proxy for the full galactic luminosity. The summed values are consistent with the full X-ray luminosity of the galaxies, but is found to have similar or narrower error ranges. The probability of a source is calculated using a Poisson distribution and the raw counts for that source. The mean of the distribution is calculated by taking the background count rate per $\rm pixel^2$ (for that specific galaxy) multiplied by the size of the {\tt srcflux} analysis region (8.875 $\rm pixel^2$), then multiplying that value by the number of those {\tt srcflux} analysis sized regions that could fit in that particular galaxy's d25 ellipse (without overlap). The distribution is then integrated over the range extending past the raw count total for the source, determining the probability that it is fluctuations in the background. This Poisson error analysis eliminated weak potential sources in galaxies 1 and 7 due to being fairly insignificant versus the background of the respective galaxy.

In the case of GALEX J02535.8$-$001357 (galaxy 3), {\tt wavdetect} found a large region of extended emission. A circular aperture of a size matching the area used by {\tt srcflux} for point sources was manually moved around the region of extended emission to find areas of high counts. The coordinates found for these high count areas were then run as point sources in {\tt srcflux}. Two such regions were identified. The third region in GALEX J02535.8$-$001357 is simply a discrete point source separated from the area of extended emission. The sum of the sources for galaxy 3 in Table 3 is the combination of the extended region and the separated source. The arrangement of the galaxy D25, extended emission region, and three sources can be seen in figure 1. The source in GALEX J100712.2$+$065736 (galaxy 7) was also found to be slightly extended, and the same procedure outlined above for GALEX J025325.8$-$001357 was followed to find a central coordinate corresponding to maximum counts. This slight adjustment to the central coordinate for the extended region only resulted in a single additional count for the source when run as a point source in {\tt srcflux}, versus running {\tt srcflux} with the central coordinate of the extended region. The second source in this galaxy is separated from the area of extended emission.

 \begin{figure*}
   \includegraphics[width=2\columnwidth]{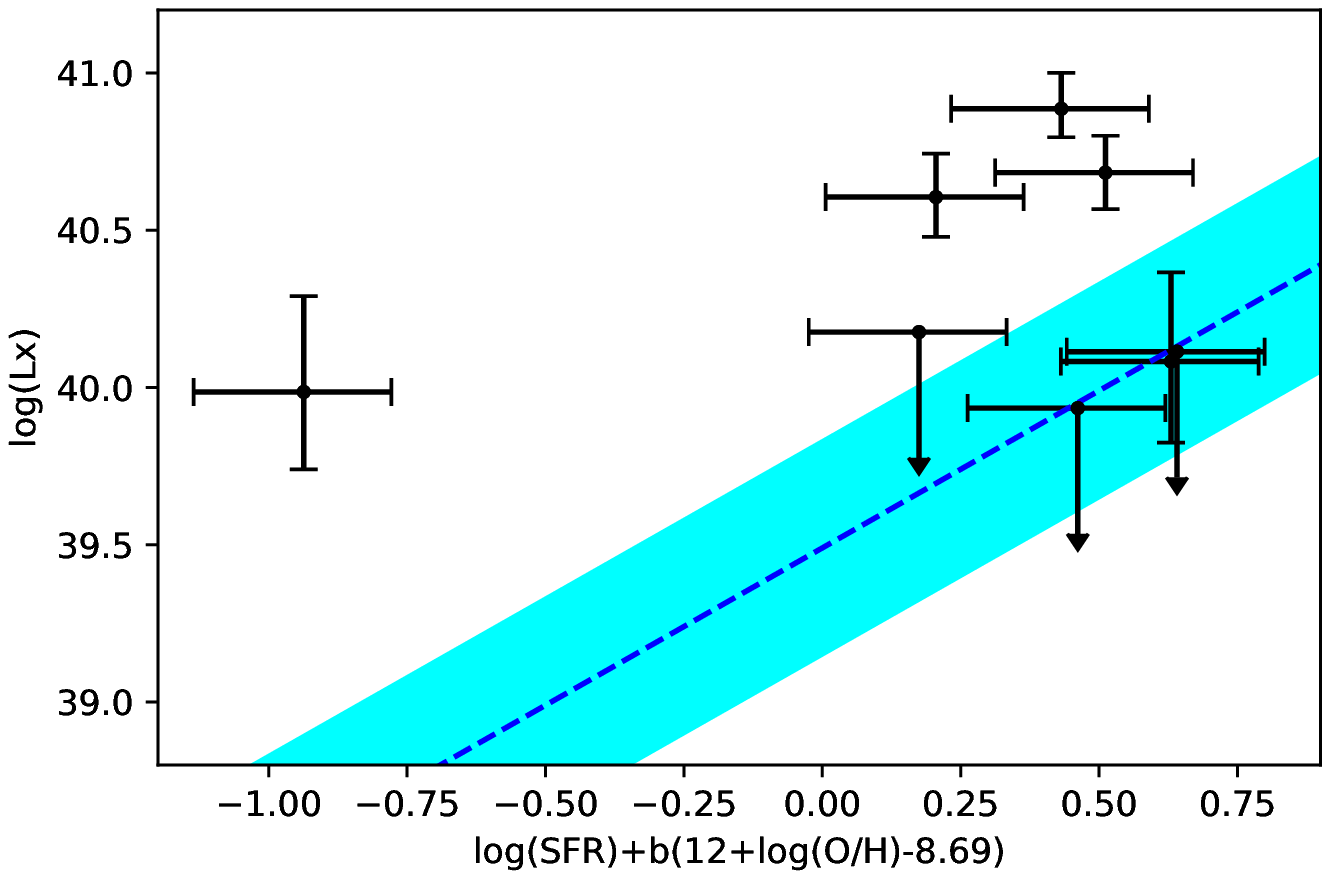}
   \caption{$\rm L_x$, SFR, Metallicity relation from \citet{Brorby2016}, with the galaxy sample from this paper plotted on top. Shaded region is the 0.34 dex dispersion around the best fit line. Galaxies with only upper limits are plotted with a downward arrow. Horizontal error is a combination of star formation rate error (assumed 30\%) and metallicity error, and does not include error in the parameters b and c of the relation equation.}
\end{figure*}

 \begin{figure*}
   \includegraphics[width=2\columnwidth]{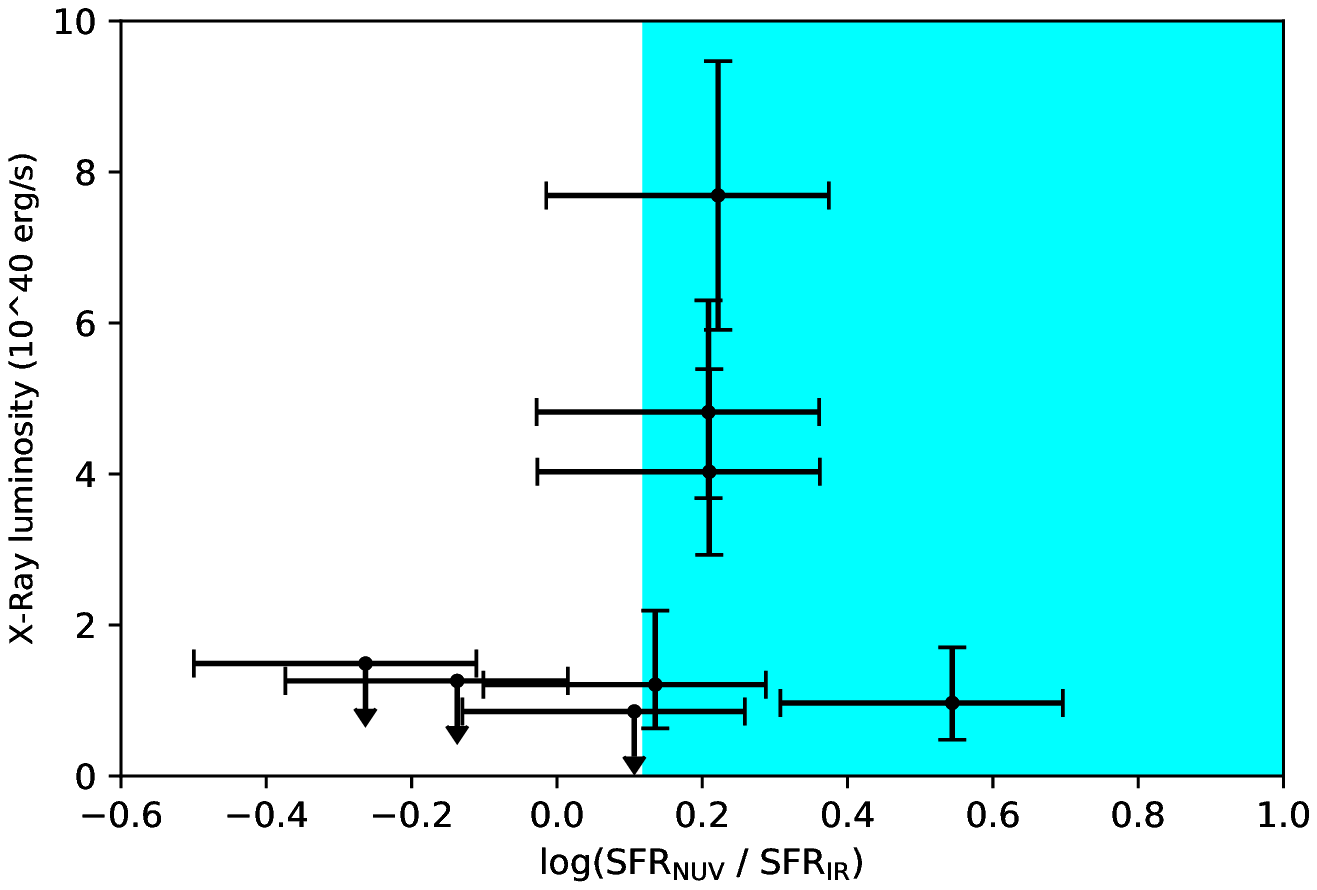}
   \caption{Comparison of $\rm SFR_{NUV}/SFR_{IR}$ versus X-ray luminosity for the 8 galaxies in the \citet{Bergvall2013} sample. The 5 galaxies with detected X-ray sources all fall within the shaded region. The edge of the shaded region is at $\rm SFR_{NUV}/SFR_{IR} = 1.33$, halfway between the neighboring data points (one with an X-ray source, one without).}
\end{figure*}

\section{Results and Discussion}\label{conc}
\label{sec:con} 

Eight sources were successfully detected in our sample of eight potential Lyman continuum emitting galaxies. Three of the eight galaxies were not detected to any significance in X-rays. Two galaxies had multiple sources. The X-ray luminosities of these sources were found to be $5 \times 10^{39}$ erg $ \rm s^{-1}$ or greater, with the brightest source detected having a luminosity of $7.7 \times 10^{40}$ erg $ \rm s^{-1}$. The galaxies in this sample are too distant to determine if these sources are individual binaries or ensembles of binaries. However, comparisons between measured and expected X-ray luminosities can serve as evidence of an enhanced population of XRBs in the sample, as most of a galaxy's X-ray luminosity is from XRBs.

Table 3 includes X-ray luminosities estimated using equation 22 from \citet{Mineo2012} (with $\rm L_{XRB}$ in $\rm erg$ $\rm s^{-1}$ and SFR in solar masses per year):
\begin{equation}
\rm L_{XRB} = 2.61\times10^{39} \, SFR
\end{equation}
The data from \citet{Mineo2012} has a dispersion of 0.34 dex around this relation. The total X-ray luminosity of the detected galaxies in our sample lie above this relation, with an average offset of 0.89 dex and the furthest offset (galaxy 8) being 1.31 dex  above the estimated value. The upper limit for X-ray luminosity of galaxy 4 also lies above the relation and deeper observations are needed. The upper limits for galaxies 1 and 6 lie within the relation.

This apparent enhancement can be further studied by including the effects of metallicity. Figure 3 compares the X-ray luminosities and star formation rates of these eight galaxies to the $\rm L_X$-SFR-metallicity relationship described in \citet{Brorby2016}, using metallicities from Table 1. Galaxy 2 has a metallicity outside of the range studied in \citet{Brorby2016}, but is included for completeness of the sample. The formula for the relation is as follows, with $\rm L_X$ (X-ray luminosity) in $\rm erg$ $\rm s^{-1}$ and SFR in solar masses per year:
\begin{equation}
\rm log(L_X) = a \, log(SFR) + b \, log((O/H)/(O/H)_{\sun}) +c
\end{equation}
In accordance with \citet{Brorby2016}, parameters a, b, and c are set to 1, $\rm -0.59 \pm 0.13$, and $\rm 39.49 \pm 0.09$, respectively. Four of the galaxies with X-ray detections are elevated above the plane described in that paper. The entire sample has a reduced $\rm \chi^2$ of 2.28, corresponding to 98\% significant deviation from the \citet{Brorby2016} relation. The galaxies with only upper limits on luminosity are treated as being consistent with the relation in the $\rm \chi^2$ calculation.

If these galaxies are leaking Lyman continuum radiation, this excess of bright X-ray sources could have helped by clearing channels out of the galaxies. This ties nicely into the weak emission lines, since the galaxies featuring these detected X-ray binaries could then be depleted in dust from the binary winds, allowing for extensive ionization to have occurred. \Citet{Bergvall2013} found estimated starburst ages of approximately 20 million years for all eight galaxies using $\rm H_\beta$ equivalent width. This corresponds to a period of time after the end of a starburst where HMXRBs are expected to be active. If these galaxies are indeed found to be Lyman leaking galaxies, then these results can also further serve to solidify the relationship between X-ray binaries and Lyman leakage.

In comparison, the starburst ages commonly reported for known Lyman leakers Haro 11 and Tololo 1247-232 are a bit less, around 1 million years for Haro 11 \Citep{Keenan2017,Adamo2010}, and less than 4 million years for Tololo 1247-232 \Citep{Rosa2007}. Note that because this time frame is so short, only the most massive stars have evolved into HMXRBs. Haro 11 and Tololo 1247-232 further differ from our sample in that both are within the $\rm L_X$-SFR-metallicity trend from \Citet{Brorby2016}.They also feature much higher SFRs, with Haro 11 in particular having a SFR of 98.1 \citep{Kaaret2017}.

Another possible linkage between the amount of dust present and XRBs can be seen in Figure 4. The figure shows a plot of the ratio of NUV star formation rate to IR star formation rate, versus total X-ray luminosity for each galaxy. There are a few apparent trends to take note of in this plot. First is that X-ray luminosity tends to increase as the star formation rate's UV component becomes increasingly dominant. The most notable outlying data point to this trend is the point on the far right, which corresponds to galaxy 2, the only galaxy in the sample with a significantly higher metallicity of 10.2 versus the 8.0-8.5 of the other 7 galaxies. The more compelling trend is that all galaxies in the shaded region (above a ratio of roughly 1.3) feature X-ray sources and the galaxies below that ratio feature none. This result is found to be 99.5\% significant by a binomial distribution analysis, and may represent the difference between star formation regions shrouded in dust versus those that are not. This may in turn further highlight the importance of X-ray binaries in enabling leakage from galaxies, if the presence of these detected X-ray sources are also responsible for the lack of dust present in these star formation regions. Future observations could be done to study this trend in depth. If this trend is found to hold true, then of the ratio of NUV SFR to IR SFR may serve as a tool to identify Lyman continuum leakers.

\section*{Acknowledgements}
Support for this work was provided by the National Aeronautics and Space Administration through Chandra Award Number GO5-18076X and GO5-16081X issued by the Chandra X-ray Center, which is operated by the Smithsonian Astrophysical Observatory for and on behalf of the National Aeronautics Space Administration under contract NAS8-03060.

We acknowledge the usage of the HyperLeda database (http://leda.univ-lyon1.fr).

%%%%%%%%%%%%%%%%%%%% REFERENCES %%%%%%%%%%%%%%%%%%

%%%%%%%%%%%%%%%%%%%%%%%%%%%%%%%%%%%%%%%%%%%%%%%%%%

%%%%%%%%%%%%%%%%% APPENDICES %%%%%%%%%%%%%%%%%%%%%

%\appendix

%\section{Some extra material}

%%%%%%%%%%%%%%%%%%%%%%%%%%%%%%%%%%%%%%%%%%%%%%%%%%

% Don't change these lines
\bsp	% typesetting comment
\label{lastpage}
\end{document}